\documentclass{revtex4}
\usepackage{graphicx}
\usepackage{fancyhdr}
\usepackage{amsmath}
\usepackage{color}

\begin{document}

\title{Testability of the Higgs inflation scenario in a radiative seesaw model
\footnote{This proceeding paper is based on Ref.~\cite{Kanemura:2012ha}.}
}


\author{Toshinori Matsui}
\affiliation{Department of Physics, University of Toyama, Toyama 930-8555, Japan}

\begin{abstract}
The Higgs inflation scenario is an approach to realize the inflation, in which the Higgs boson plays a role of the inflaton without introducing a new particle. 
We investigate a Higgs inflation scenario in the so-called radiative seesaw model proposed by E.~Ma. 
We find that a part of  parameter regions where additional scalar fields can play a role of an inflaton is compatible with the current LHC results, the current data from neutrino experiments and those of the dark matter abundance as well as the direct search. 
We show that we can partially test this model by measuring masses of scalar bosons at the International Linear Collider. 
\end{abstract}

\maketitle

\thispagestyle{fancy}


%%%%%%%%%%%%%%%%%%%%%%%%%%%%%
%%%%%%%% Introduction %%%%%%%
%%%%%%%%%%%%%%%%%%%%%%%%%%%%%
\section{Introduction}

%%Higgs inflation
In 2012, the LHC discovered a new particle with the mass of 126~GeV~\cite{atlas, cms}. 
The particle is regarded as the Higgs boson predicted in the Standard Model (SM) of elementary particles. 
The discovery of the Higgs boson means that all the particle contents in the SM are completed. 
The LHC is now searching for indications of new physics, and is trying to measure the deviation in the coupling from the SM. 
On the other hand, the cosmic observations such as the experiments at WMAP and Planck have reported the new results~\cite{WMAP, Planck}. 
These experiments measure the temperature fluctuation of the cosmic microwave background precisely, by which we can impose constraints on the models of inflation. 
Cosmic inflation at the early Universe~\cite{inf}, which is a promising candidate to solve cosmological problems such as the horizon problem and the flatness problem, requires an additional scalar boson, the inflaton. 
We consider the Higgs inflation scenario where the Higgs boson plays a role of the inflaton. 
In the minimal model of this scenario~\cite{Hinf}, we do not have to introduce any other particle in addition to the particle contents in the SM to explain an inflation.  

%%Vacuum stability
However, it would be difficult to realize the Higgs inflation scenario in the minimal model.
Assuming the SM with one Higgs doublet, the vacuum stability argument indicates that the model can be well defined only below the energy scale where the running coupling of the Higgs self-coupling becomes zero. 
For the Higgs boson mass to be 126~GeV with the top quark mass to be 173.1~GeV and for the coupling for the strong force to be $\alpha_s =$ 0.1184, the critical energy scale is estimated to be around $10^{10}$~GeV using the NNLO calculation, although the uncertainty due to the values of the top quark mass and $\alpha_s$ is not small~\cite{lambda_run}. 
The vacuum seems to be metastable when we assume that the model holds up to the Planck scale. 
This kind of analysis gives a strong constraint on the scenario of the Higgs inflation, because the inflation occurs at the energy scale where the vacuum stability is not guaranteed in the SM. 
Recently, a viable model for the Higgs inflation has been proposed, in which the Higgs sector is extended including an additional scalar field~\cite{other_Hinf, 2Hinf}. 
There is also another problem in the minimal model, which comes from unitarity argument~\cite{uni_break,uni_care}.

%%New physics and radiative seesaw
Extending the Higgs sector from the SM one, we may expect to reveal new physics that can explain phenomena such as neutrino oscillation, existence of dark matter and baryon asymmetry of the Universe. 
Here, we extend the Higgs inflation model in the framework of a radiative seesaw scenario by E.~Ma~\cite{Kanemura:2012ha}. 
The radiative seesaw scenario is a way to explain tiny neutrino masses, where they are radiatively induced at the loop level by introducing $Z_2$-odd scalar fields and $Z_2$-odd right-handed neutrinos~\cite{KNT,Ma,AKS}. 
An interesting characteristic feature in these radiative seesaw models is that dark matter candidates automatically enter into the model because of the $Z_2$ parity. 

%%In this talk
In this work, we discuss a simple model to explain inflation, neutrino masses and dark matter simultaneously, which is based on the simplest radiative seesaw model~\cite{Ma}. 
Both the Higgs boson and neutral components of the $Z_2$-odd scalar doublet can satisfy conditions on the slow-roll inflation~\cite{slow-roll} and vacuum stability up to the inflation scale. 
We find that a part of the parameter region where these scalar fields can play a role of the inflaton 
is compatible with the current LHC results, the current data from neutrino experiments and 
those of the dark matter abundance as well as the direct search~\cite{XENON100}. 
A phenomenological consequence of scenario results in a specific mass spectrum of scalar fields, 
which can be tested at the International Linear Collider (ILC)~\cite{ILC1}.

%%%%%%%%%%%%%%%%%%%%%%%%%%%%%
%%%%%%%%%%% Model %%%%%%%%%%%
%%%%%%%%%%%%%%%%%%%%%%%%%%%%%
\section{Extension to a radiative seesaw model}

%%Symmetry and coupling constants
We extend the Higgs inflation model in the framework of a radiative seesaw scenario~\cite{Ma}. 
In this model, there are the $Z_2$-odd scalar doublet field $\Phi_2$ and right-handed neutrino $\nu_R^{}$ in addition to the  $Z_2$-even SM Higgs doublet field $\Phi_1$ due to the invariance under the unbroken discrete $Z_2$ symmetry~\cite{Ma}. 
Because Dirac Yukawa couplings of neutrinos are forbidden by the $Z_2$ symmetry, the Yukawa interaction for leptons is given by ${\cal L}_{Yukawa} = Y_\ell \overline{L_L}\Phi_1\ell_R+Y_\nu\overline{L_L}\Phi_2^c\nu_R+h.c.$ (the superscript $c$ denotes the charge conjugation). 
The scalar potential is given by~\cite{2Hinf} 
\begin{eqnarray}
V &=&\frac{M_P^2 R}{2}+(\xi_1|\Phi_1|^2 +\xi_2 |\Phi_2|^2)R +\mu_1^2 |\Phi_1|^2 + \mu_2^2 |\Phi_2|^2 
\nonumber\\
  &&+ \frac{1}{2} \lambda_1 |\Phi_1|^4 + \frac{1}{2} \lambda_2 |\Phi_2|^4 + \lambda_3|\Phi_1|^2|\Phi_2|^2 
+ \lambda_4 (\Phi_1^{\dagger} \Phi_2) (\Phi_2^{\dagger} \Phi_1) 
+ \frac{1}{2}\lambda_5((\Phi_1^{\dagger} \Phi_2)^2+h.c.), 
\label{eq:potential}
\end{eqnarray}
where $M_P$($\simeq 10^{19}$~GeV) is the Planck scale, and $R$ is the Ricci scalar.
Then, these quartic coupling constants should satisfy the following constraints on the unbounded-from-below conditions at the tree level; 
\begin{eqnarray}
\lambda_1>0,\ \ \lambda_2>0,\ \ \lambda_3+\lambda_4+\lambda_5+\sqrt{\lambda_1 \lambda_2}>0, 
 \label{eq:vs}
\end{eqnarray}
and we impose the conditions of triviality; 
\begin{eqnarray}
\lambda_i \lesssim 2\pi. 
\label{eq:tri}
\end{eqnarray}

%%VEV and scalar boson mass
Assuming $\mu_1^2 < $0 and $\mu_2^2 >$ 0, $\Phi_1$ obtains the vacuum expectation value (VEV) $v$ ($=\sqrt{-2\mu_1^2/\lambda_1}$), while $\Phi_2$ cannot get the VEV because of the unbroken $Z_2$ symmetry. 
The lightest $Z_2$-odd particle is stabilized by the $Z_2$ parity, and it can act as the dark matter as long as it is electrically neutral. 
Mass eigenstates of the scalar bosons are the SM-like $Z_2$-even Higgs scalar boson ($h$), the $Z_2$-odd CP-even scalar boson ($H$), the $Z_2$-odd CP-odd scalar boson ($A$) and $Z_2$-odd charged scalar bosons ($H^\pm$). 
Masses of these scalar bosons are given by~\cite{Ma}; $m_h^2=\lambda_1 v^2, \ m_H^2=\mu_2^2 +\frac{1}{2}(\lambda_3+\lambda_4+\lambda_5) v^2, \ m_A^2=\mu_2^2 +\frac{1}{2}(\lambda_3+\lambda_4-\lambda_5) v^2, \ m_{H^{\pm}}^2=\mu_2^2 +\frac{1}{2}\lambda_3 v^2$.

%%%%%%%%%%%%%%%%%%%%%%%%%%%%%%%%%%
%%%   Constraint   %%%
%%%%%%%%%%%%%%%%%%%%%%%%%%%%%%%%%%
\section{Constraints on the parameters}

%%%%%%%%%%%%%%%%%%%%%%%%%%%%%%%%%%
%\subsection{Introduction}
For the Higgs inflation scenario in our model defined in the previous section, there are nine parameters in the scalar sector; i.e., $\xi_1$, $\xi_2$, $\mu_1^2$, $\mu_2^2$, $\lambda_1$, $\lambda_2$, $\lambda_3$, $\lambda_4$ and $\lambda_5$. 
They must satisfy the vacuum stability condition on the running of the scalar coupling constants and the constraint from the slow-roll inflation, the dark matter data and the neutrino data. 
We find that a part of  parameter regions is compatible with all constraints. 
Then, we can get the possible mass spectrum for additional scalar bosons in our model~\cite{Kanemura:2012ha}. 

%%%%%%%%%%%%%%%%%%%%%%%%%%%%%%%%%%
%\subsection{Inflation}
First, we discuss the constraint from the slow-roll inflation.  
In order that some of the scalar bosons play a role of the inflaton, 
we need to impose following conditions~\cite{2Hinf}; 
\begin{eqnarray}
\lambda_2\xi_1-(\lambda_3+\lambda_4)\xi_2&>&0,
\nonumber\\
\lambda_1\xi_2-(\lambda_3+\lambda_4)\xi_1&>&0,
\nonumber\\
\lambda_1\lambda_2-(\lambda_3+\lambda_4)^2&>&0.
 \label{eq:vs2}
\end{eqnarray}
Parameters in the scalar potential should satisfy 
the constraint from the power spectrum~\cite{WMAP, 2Hinf};
\begin{eqnarray}
\xi_2 \sqrt{\frac{2(\lambda_1+a^2\lambda_2-2a(\lambda_3+\lambda_4))}{\lambda_1\lambda_2-(\lambda_3+\lambda_4)^2}} \simeq 5\times 10^{4}, \ \ \ \ \ 
\frac{\lambda_5}{\xi_2} 
\frac{a\lambda_2 - (\lambda_3+\lambda_4)}{\lambda_1+a^2\lambda_2-2a(\lambda_3+\lambda_4)}
\lesssim 4\times 10^{-12},
 \label{eq:l5}
\end{eqnarray}
where $a$ is given as $a\equiv\xi_1/\xi_2$.
When the scalar potential satisfies the conditions in Eqs. (\ref{eq:vs2}) and (\ref{eq:l5}), 
the model could realize the inflation. 

%%%%%%%%%%%%%%%%%%%%%%%%%%%%%%%%%%
%\subsection{Dark Matter}
Second, we discuss the constraint from dark matter. 
We here assume that the CP-odd boson $A$ is the dark matter (the lightest $Z_2$-odd particle). 
When $\lambda_5$ is very small such as ${\cal O}(10^{-7})$, $A$ is difficult to act as the dark matter because the scattering process $AN\to HN$ ($N$ is a nucleon) opens and the cross section cannot be consistent with the current direct search results for dark matter~\cite{direct_Z, Kashiwase:2012xd, LopezHonorez:2006gr}. 
To avoid the process $AN\to HN$ kinematically, we here take $\lambda_5\simeq 10^{-6}$ and 
\begin{eqnarray}
a\lambda_2 - (\lambda_3+\lambda_4)\simeq 10^{-1}
\label{eq:FT}
\end{eqnarray}
at the inflation scale. 
With this choice, masses of $A$ and $H$ are almost the same value.
The co-annihilation process $AH\to XX$ via the $Z$ boson is important to explain the abundance of the dark matter where $X$ is a  particle in the SM, because the pair annihilation process $AA\to XX$ via the $h$ boson is suppressed due to the constraint from the inflation. 
Because the cross section of $AH\to XX$ depends only on the mass of the dark matter, the mass of the dark matter $A$ is constrained from the abundance of the dark matter as $128~{\rm GeV}\leq m_A\leq138~{\rm GeV}$, where we have used the nine years WMAP data~\cite{WMAP}.

%%%%%%%%%%%%%%%%%%%%%%%%%%%%%%%%%%
%\subsection{Tiny Neutrino Masses}
Third, we can explain tiny neutrino masses in this model which are generated by the one loop diagram~\cite{Ma}. 
The neutrino mass is related to $\lambda_5$ and masses of scalar bosons ($m_H^{}$ and $m_A^{}$),  which are constrained from the inflation and the dark matter. 
From the relation $(Y_\nu)_i^k(Y_\nu)_j^k/M_R^k\simeq {\cal O} (10^{-11})$~GeV$^{-1}$ where $M_R^k$ is the Majorana mass of $\nu_R^k$ ($k$=1-3) and $(Y_\nu)_i^k$ is neutrino Yukawa coupling constant, the magnitude of tiny neutrino masses can be explained. 
For example, when $M_R^k$ is ${\cal O}(1)$~TeV, $(Y_\nu)_i^k$ is ${\cal O}(10^{-2})$.

%%%%%%%%%%%%%%%%%%%%%%%%%%%%%%%%%%
%\subsection{Running of Scalar Coupling Constants}
\begin{figure}[t]
\begin{minipage}{0.497\textwidth}
 \begin{center}
  \scalebox{0.377}{\includegraphics{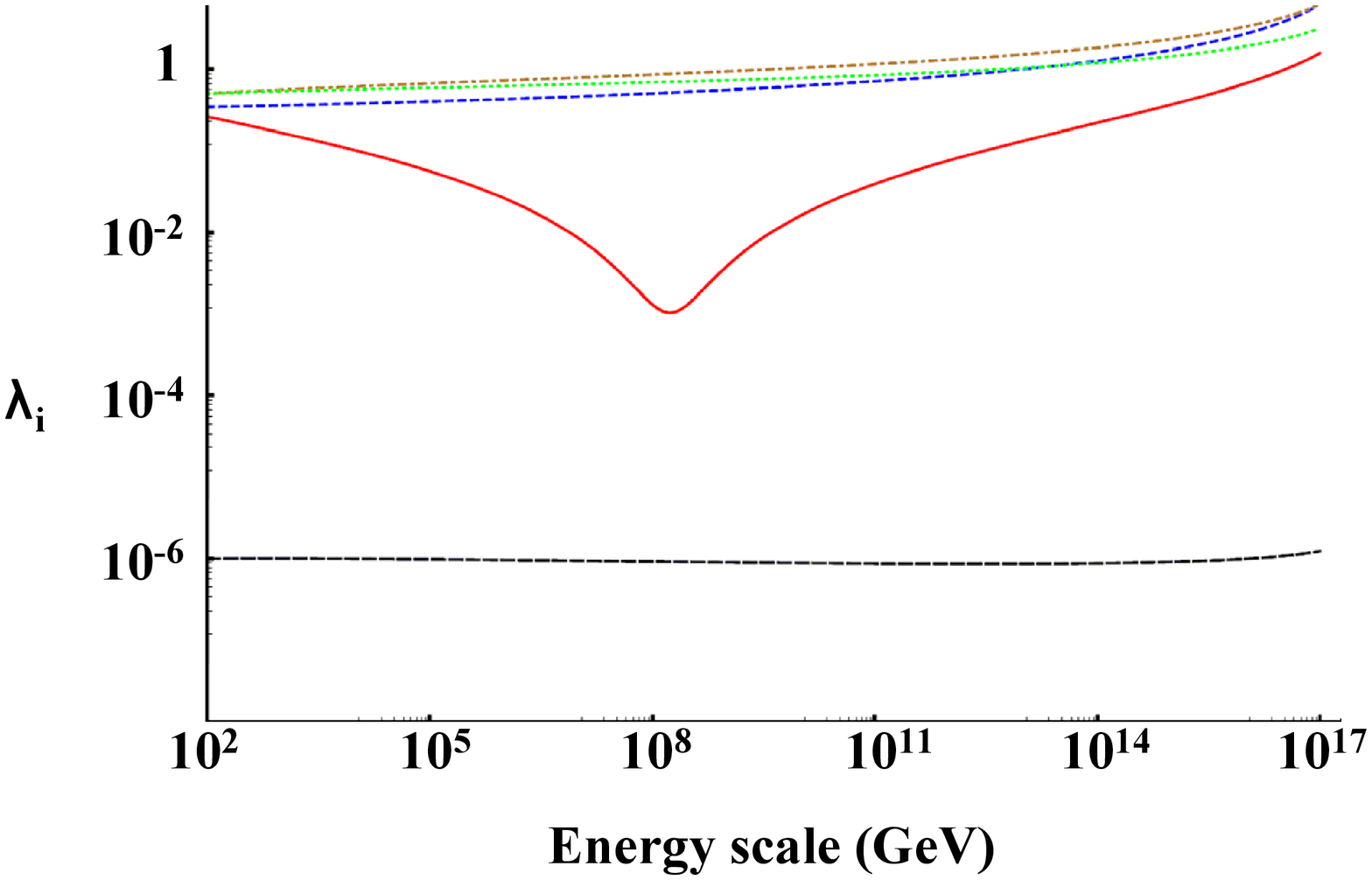}}
 \end{center}
  \end{minipage}
 \begin{minipage}{0.497\textwidth}
     \caption{Running of the scalar coupling constants.
  Red (solid), blue (dashed), brown (dot-dashed), green (dotted) and black (long-dashed) curves show 
  $\lambda_1$, $\lambda_2$, $\lambda_3$, $-\lambda_4$ and $\lambda_5$, respectively. \\}
    \label{fig:run}
\begin{center}
\makeatletter
\def\@captype{table}
\makeatother
\begin{tabular}{|c|c|c|c|c|c|}
\hline
&$\lambda_{1}$&$\lambda_{2}$&$\lambda_{3}$&$\lambda_{4}$&$\lambda_{5}$\\
\hline
 $10^{2}$~GeV
  &0.26 &0.35&0.51&-0.51&1.0$\times10^{-6}$\\
\hline
 $10^{17}$~GeV
  &1.6 &6.3&6.3&-3.2&1.2$\times10^{-6}$\\
\hline
\end{tabular}
\caption{The possible parameter set which satisfies constraints from the inflation and the dark matter at the scales of ${\cal O}(10^{2})$~GeV and ${\cal O}(10^{17})$~GeV. }
\label{table:lambda}
\end{center} 
  \end{minipage}
\end{figure}

Finally, we calculate the running of the coupling constants using the renormalization group equations~\cite{beta}. 
As shown in Fig~\ref{fig:run}, for the contribution of additional scalar bosons, this model can be stable up to the inflation scale from the electroweak scale~\cite{extended_run}. 
As numerical input parameters, we take the VEV ($v = 246~$GeV), SM-like Higgs mass ($m_h = 126$~GeV) and the allowed value for the dark matter mass ($m_A^{} = 130$~GeV). 
Further numerical input parameter comes from the perturbativity of $\lambda_2$ up to the inflation scale; i.e., $\lambda_2(\mu_{\rm inf}) = 2\pi$, where $\mu_{\rm inf}$ is the inflation scale $10^{17}$~GeV. 
The parameter set in Table~\ref{table:lambda} can be consistent with these numerical inputs and the constraints are given in Eqs.~(\ref{eq:vs})-(\ref{eq:FT}). 
Consequently, we can obtain the mass spectrum of the scalar bosons in our model as 
\begin{eqnarray}
m_h\simeq126~{\rm GeV}, \ \ \ 
m_{H^{\pm}}\simeq173~{\rm GeV},\ \ \ 
m_H\simeq130~{\rm GeV},\ \ \ 
m_A\simeq130~{\rm GeV},
 \label{eq:hmassD}
\end{eqnarray}
where the mass difference between $A$ and $H$ is about 500~KeV.
The mass spectrum is not largely changed even if $m_A^{}$ is varied with in its allowed region. 
In the next section, we consider the constraints on our model from the existing experiments and the way to test the characteristic mass spectrum in this model at the future collider experiment.

%%%%%%%%%%%%%%%%%%%%%%%%%%%%%%%%%%
%%%   Phenomenology   %%%
%%%%%%%%%%%%%%%%%%%%%%%%%%%%%%%%%%
\section{Phenomenology}

%LEP
The LEP experiment constrains masses of the $Z_2$-odd scalar bosons.
The mass of charged scalar bosons $m_{H^\pm}^{}$ should be lager than 70-90~GeV by the LEP~\cite{LEP_direct, LEP_pm}.
This constraint is satisfied in our model ($m_{H^{\pm}}\simeq173~{\rm GeV}$).
Furthermore, $m_{H}^{} + m_{A}^{}$ should be larger than $m_Z^{}$, and the combination of $m_{H}^{}$ and $m_{A}^{}$ is bounded by $H$$A$ production by the LEP date~\cite{LEP_direct, LEP_HApair}. 
However, when $m_{H}^{} - m_{A}^{} < 8$~GeV, masses of neutral $Z_2$-odd scalar boson loop diagrams are not really constrained by the LEP~\cite{LEP_direct, LEP_HApair}. 
On the other hand, the contributions to the electroweak parameters~\cite{STdef} from additional scalar bosons loops which are given by~\cite{ST1, ST2} are also consistent with the electroweak precision data with 90\% Confidence Level (C.L.)~\cite{ST2}.

%LHC
Next, we consider the way to test at the LHC.
According to Refs.~\cite{LHC1, LHC2, LHC3}, they conclude that it could be difficult to test $pp\to AH^+/HH^+/H^+H^-$ processes because the cross sections of the background processes are very large. 
The process of $pp\to AH$ could be tested with about the 3$\sigma$ C.L. with the various benchmark points for $m_A$ and $m_H$.
However, it would be difficult to test $pp\to AH$ in our scenario, because $m_H$ and $m_A$ are almost degenerate in our scenario, and the event number of $pp\to AH$ is negligibly small after imposing the basic cuts~\cite{LHC1, LHC2, LHC3}. 
Furthermore, as the total decay width of $H$ is about $10^{-29}$~GeV, $H$ would pass through the detector. 
Therefore, this signal is also difficult to be detected at the LHC. 

\begin{figure}[t]
 \begin{center}
  \scalebox{0.377}{\includegraphics[angle=-90]{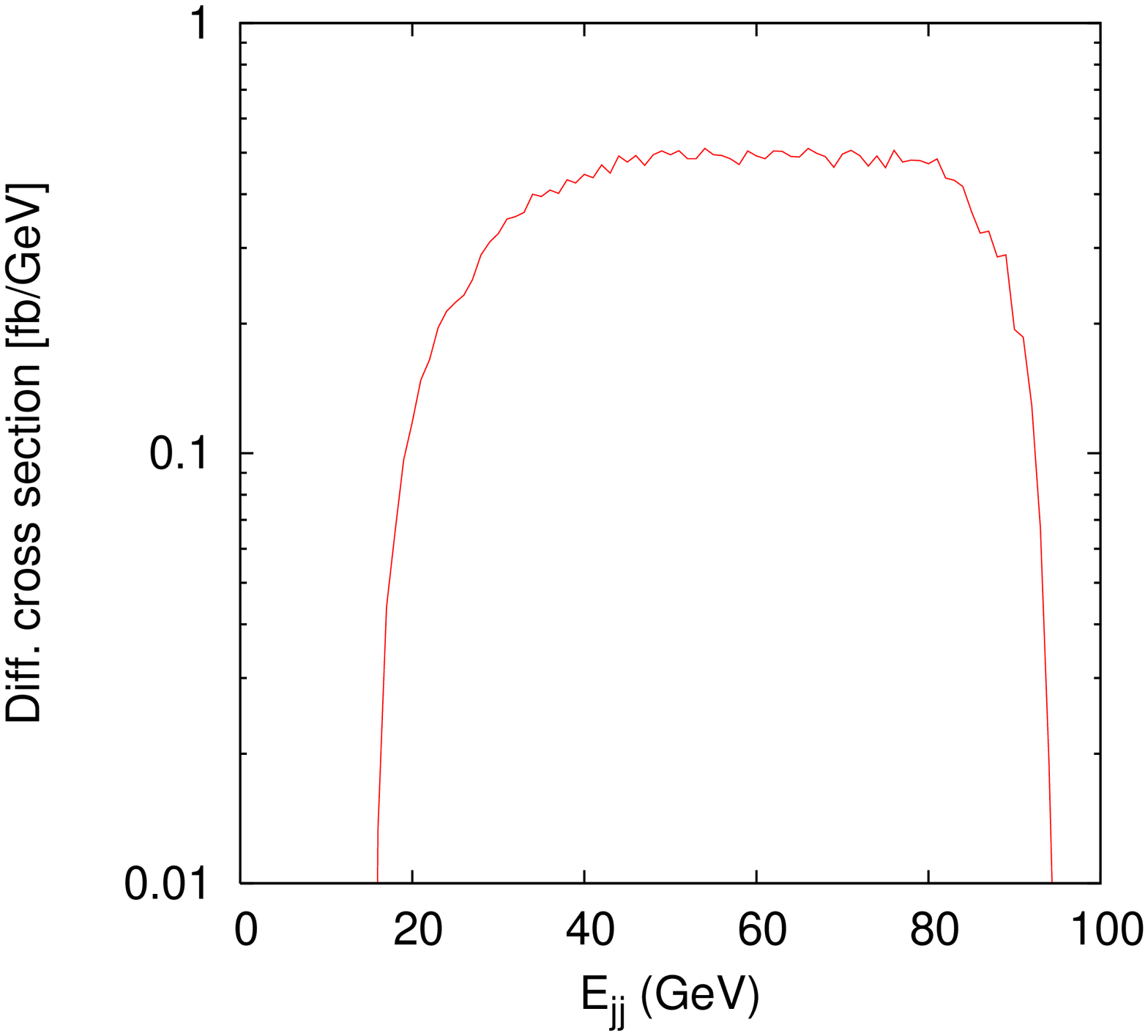}}
  \caption{The distribution of $E_{jj}$ for the differential cross section for $e^+e^-\to H^+H^-\to W^{+(*)}W^{-(*)}AA\to jj\ell\nu AA$. In our parameter set, the endpoint of $E_{jj}$ is estimated at $15~{\rm GeV} < E_{jj} < 94~{\rm GeV}$.}
  \label{fig:Ejj}
 \end{center}
\end{figure}

%ILC-1
Finally, we discuss the signals of $H, A$ and $H^\pm$ at the ILC with $\sqrt{s}=500$~GeV. 
In the following, we use Calchep~2.5.6 for numerical evaluation~\cite{calc}.
We focus on the $H^\pm$ pair production process: $e^+e^-\to Z^*(\gamma^*)\to H^+H^-\to W^{+(*)}W^{-(*)}AA\to jj\ell\nu AA$ ($j$ denotes a hadron jet)~\cite{ILC2}. 
Because of the kinematical reason, the energy of the two-jet system $E_{jj}$ satisfies the following equation; 
\begin{eqnarray}
\frac{m_{H^{\pm}}^2-m_A^2}{\sqrt{s}+2\sqrt{s/4-m_{H^{\pm}}^2}} < E_{jj} < \frac{m_{H^{\pm}}^2-m_A^2}{\sqrt{s}-2\sqrt{s/4-m_{H^{\pm}}^2}}. 
 \label{eq:Ejj}
\end{eqnarray}
In our parameter set, the distribution of $E_{jj}$ for the differential cross section in this process is shown in Fig.~\ref{fig:Ejj}. 
The important background processes against this process, which are $e^+e^-\to W^{+}W^{-}\to jj\ell\overline{\nu}$ and $e^+e^-\to Z(\gamma )Z\to jj\ell\overline{\ell}$ with a missing $\overline{\ell}$ event, could be well reduced by imposing an appropriate kinematic cuts. 
Then, we expect that $m_{H^\pm}$ and $m_A$ can be measured by using the endpoints of $E_{jj}$ at the ILC after the background reduction. 

On the other hand, we consider $HA$ production: $e^+e^-\to Z^*\to HA\to AAZ^*\to AAjj$ at the ILC. 
If the mass difference between $m_A$ and $m_H$ is sizable, it could also be detected by using the endpoint of $E_{jj}$. 
However, $m_A$ and $m_H$ are almost degenerate in our scenario.
When we detect $H^\pm$ but we cannot detect the clue of this process at the ILC, it seems that $m_A$ and $m_H$ are almost same value.

%%%%%%%%%%%%%%%%%%%%%%%%%%%%%%%%%%
%%% Conclusion %%%
%%%%%%%%%%%%%%%%%%%%%%%%%%%%%%%%%%
\section{Conclusion}
We have studied the Higgs inflation model in the framework of a radiative seesaw scenario. 
In our model, we may be able to explain inflation, neutrino masses and dark matter simultaneously. 
We find that a part of parameter regions is compatible with all constraints which come from the conditions of the slow-roll inflation, the current LHC results, the current data from neutrino experiments and those of the dark matter abundance as well as the direct search results. 
We can test this scenario by measuring masses of scalar bosons at the ILC with $\sqrt{s}=500$~GeV.

\begin{acknowledgments}
This work is collaboration with Shinya Kanemura and Takehiro Nabeshima. 
I would like to thank them for their support.
\end{acknowledgments}

\bigskip % extra skip inserted

\begin{thebibliography}{99} % Use for 10-99 references
%\cite{Kanemura:2012ha}
\bibitem{Kanemura:2012ha}
  S.~Kanemura, T.~Matsui, T.~Nabeshima and ,
  %``Higgs inflation in a radiative seesaw model,''
  arXiv:1211.4448 [hep-ph].
  %%CITATION = ARXIV:1211.4448;%%
  %1 citations counted in INSPIRE as of 01 Apr 2013

%%%%%%%%%%  atlas  %%%%%%%%%%%%%%
\bibitem{atlas}
%\cite{:2012gk}
%\bibitem{:2012gk} 
  G.~Aad {\it et al.}  [ATLAS Collaboration],
  %``Observation of a new particle in the search for the Standard Model Higgs boson with the ATLAS detector at the LHC,''  
  Phys.\ Lett.\ B {\bf 716} (2012) 1.  %[arXiv:1207.7214 [hep-ex]].  %%CITATION = ARXIV:1207.7214;%%

%%%%%%%%%%  cms  %%%%%%%%%%%%%%
\bibitem{cms} 
%\cite{:2012gu}
%\bibitem{:2012gu} 
  S.~Chatrchyan {\it et al.}  [CMS Collaboration],
  %``Observation of a new boson at a mass of 125 GeV with the CMS experiment at the LHC,''  %Submitted to: Phys.Lett.B 
   Phys.\ Lett.\ B {\bf 716} (2012) 30.
   %arXiv:1207.7235 [hep-ex].  %%CITATION = ARXIV:1207.7235;%%

%%%%%%%%%%  WMAP  %%%%%%%%%%%%%%
\bibitem{WMAP}
%\cite{Larson:2010gs}
%\bibitem{Larson:2010gs}
  D.~Larson {\it et al.},
  %``Seven-Year Wilkinson Microwave Anisotropy Probe (WMAP) Observations: Power
  %Spectra and WMAP-Derived Parameters,''
  Astrophys.\ J.\ Suppl.\  {\bf 192} (2011) 16, 
%  [arXiv:1001.4635 [astro-ph.CO]].
  %%CITATION = APJSA,192,16;%%
%\cite{Hinshaw:2012fq}
%\bibitem{Hinshaw:2012fq}
  G.~Hinshaw, D.~Larson, E.~Komatsu, D.~N.~Spergel, C.~L.~Bennett, J.~Dunkley, M.~R.~Nolta and M.~Halpern {\it et al.},
  %``Nine-Year Wilkinson Microwave Anisotropy Probe (WMAP) Observations: Cosmological Parameter Results,''  
  arXiv:1212.5226 [astro-ph.CO].  %%CITATION = ARXIV:1212.5226;%%  %98 citations counted in INSPIRE as of 11 Mar 2013

%%%%%%%%%%  Planck  %%%%%%%%%%%%%%
%\cite{Ade:2013xsa}
\bibitem{Planck}
  P.~A.~R.~Ade {\it et al.}  [ Planck Collaboration],
  %``Planck 2013 results. I. Overview of products and scientific results,''
  arXiv:1303.5062 [astro-ph.CO].
  %%CITATION = ARXIV:1303.5062;%%
  %16 citations counted in INSPIRE as of 12 Apr 2013

\bibitem{inf}
%\cite{Guth:1980zm}
%\bibitem{Guth:1980zm}
  A.~H.~Guth,
  %``The Inflationary Universe: A Possible Solution To The Horizon And Flatness
  %Problems,''
  Phys.\ Rev.\  D {\bf 23} (1981) 347;
  %%CITATION = PHRVA,D23,347;%%
%\bibitem{Sato:1980yn}
  K.~Sato,
  %``First Order Phase Transition of a Vacuum and Expansion of the Universe,''
  Mon.\ Not.\ Roy.\ Astron.\ Soc.\  {\bf 195} (1981) 467.
  %%CITATION = MNRAA,195,467;%%
%\cite{EliasMiro:2011aa}

%%\cite{Bezrukov:2007ep}
\bibitem{Hinf} 
  F.~L.~Bezrukov and M.~Shaposhnikov,
  %``The Standard Model Higgs boson as the inflaton,''  
  Phys.\ Lett.\ B {\bf 659} (2008) 703.  %[arXiv:0710.3755 [hep-th]].  %%CITATION = ARXIV:0710.3755;%%

\bibitem{lambda_run} 
%\cite{DeSimone:2008ei}
%\bibitem{DeSimone:2008ei} 
  A.~De Simone, M.~P.~Hertzberg and F.~Wilczek,
  %``Running Inflation in the Standard Model,''  
  Phys.\ Lett.\ B {\bf 678} (2009) 1;  %[arXiv:0812.4946 [hep-ph]].  %%CITATION = ARXIV:0812.4946;%%
%
%\cite{Bezrukov:2009db}
%\bibitem{Bezrukov:2009db} 
  F.~Bezrukov and M.~Shaposhnikov,
  %``Standard Model Higgs boson mass from inflation: Two loop analysis,''  
  JHEP {\bf 0907} (2009) 089;  %[arXiv:0904.1537 [hep-ph]].  %%CITATION = ARXIV:0904.1537;%%
%S
  J.~Elias-Miro, J.~R.~Espinosa, G.~F.~Giudice, G.~Isidori, A.~Riotto and A.~Strumia,
  %``Higgs mass implications on the stability of the electroweak vacuum,''  
  Phys.\ Lett.\ B {\bf 709} (2012) 222;  %[arXiv:1112.3022 [hep-ph]].  %%CITATION = ARXIV:1112.3022;%%
%\cite{Degrassi:2012ry}
%\bibitem{Degrassi:2012ry} 
  G.~Degrassi, S.~Di Vita, J.~Elias-Miro, J.~R.~Espinosa, G.~F.~Giudice, G.~Isidori and A.~Strumia,
  %``Higgs mass and vacuum stability in the Standard Model at NNLO,''  
  JHEP {\bf 1208} (2012) 098.  %[arXiv:1205.6497 [hep-ph]].  %%CITATION = ARXIV:1205.6497;%%

%\cite{Arina:2012fb}
\bibitem{other_Hinf} 
%\cite{Lerner:2009xg}
%\bibitem{Lerner:2009xg}
  R.~N.~Lerner and J.~McDonald,
  %``Gauge singlet scalar as inflaton and thermal relic dark matter,''  
  Phys.\ Rev.\ D {\bf 80} (2009) 123507,%  [arXiv:0909.0520 [hep-ph]].  %%CITATION = ARXIV:0909.0520;%%  %44 citations counted in INSPIRE as of 11 Mar 2013
%
%\cite{Lerner:2011ge}
%\bibitem{Lerner:2011ge}
  R.~N.~Lerner and J.~McDonald,
  %``Distinguishing Higgs inflation and its variants,''  
  Phys.\ Rev.\ D {\bf 83} (2011) 123522,%  [arXiv:1104.2468 [hep-ph]].  %%CITATION = ARXIV:1104.2468;%%  %9 citations counted in INSPIRE as of 11 Mar 2013
%
%\cite{Lerner:2011it}
%\bibitem{Lerner:2011it}
  R.~N.~Lerner and J.~McDonald,
  %``Unitarity-Violation in Generalized Higgs Inflation Models,''  
  JCAP {\bf 1211} (2012) 019,%  [arXiv:1112.0954 [hep-ph]].  %%CITATION = ARXIV:1112.0954;%%  %5 citations counted in INSPIRE as of 11 Mar 2013
%
  C.~Arina, J.~-O.~Gong and N.~Sahu,
  %``Unifying darko-lepto-genesis with scalar triplet inflation,''  
  Nucl.\ Phys.\ B {\bf 865} (2012) 430.  %[arXiv:1206.0009 [hep-ph]].  %%CITATION = ARXIV:1206.0009;%

%\cite{Gong:2012ri}
\bibitem{2Hinf} 
  J.~-O.~Gong, H.~M.~Lee and S.~K.~Kang,
  %``Inflation and dark matter in two Higgs doublet models,''  
  JHEP {\bf 1204} (2012) 128.  %[arXiv:1202.0288 [hep-ph]].  %%CITATION = ARXIV:1202.0288;%%
  
\bibitem{uni_break}
%\cite{Burgess:2009ea}
%\bibitem{Burgess:2009ea}
  C.~P.~Burgess, H.~M.~Lee and M.~Trott,
  %``Power-counting and the Validity of the Classical Approximation During
  %Inflation,''
  JHEP {\bf 0909} (2009) 103;
  %[arXiv:0902.4465 [hep-ph]]~;
  %%CITATION = JHEPA,0909,103;%%
%\cite{Burgess:2010zq}
%\bibitem{Burgess:2010zq}
  %C.~P.~Burgess, H.~M.~Lee and M.~Trott,
  %``Comment on Higgs Inflation and Naturalness,''
  JHEP {\bf 1007} (2010) 007;
  %[arXiv:1002.2730 [hep-ph]]~;
  %%CITATION = JHEPA,1007,007;%%
%\cite{Barbon:2009ya}
%\bibitem{Barbon:2009ya}
  J.~L.~F.~Barbon and J.~R.~Espinosa,
  %``On the Naturalness of Higgs Inflation,''
  Phys.\ Rev.\  D {\bf 79} (2009) 081302;
  %[arXiv:0903.0355 [hep-ph]]~;
  %%CITATION = PHRVA,D79,081302;%%
%\cite{Hertzberg:2010dc}
%\bibitem{Hertzberg:2010dc}
  M.~P.~Hertzberg,
  %``On Inflation with Non-minimal Coupling,''
  JHEP {\bf 1011} (2010) 023.
  %[arXiv:1002.2995 [hep-ph]].
  %%CITATION = JHEPA,1011,023;%%
  
%%\cite{Giudice:2010ka}
\bibitem{uni_care} 
  G.~F.~Giudice and H.~M.~Lee,
  %``Unitarizing Higgs Inflation,''  
  Phys.\ Lett.\ B {\bf 694} (2011) 294.  %[arXiv:1010.1417 [hep-ph]].  %%CITATION = ARXIV:1010.1417;%%
  
\bibitem{KNT}
%\cite{Krauss:2002px}
%\bibitem{Krauss:2002px}
  L.~M.~Krauss, S.~Nasri and M.~Trodden,
  %``A model for neutrino masses and dark matter,''
  Phys.\ Rev.\  D {\bf 67} (2003) 085002;
%  [arXiv:hep-ph/0210389].
  %%CITATION = PHRVA,D67,085002;%%
%
%\cite{Cheung:2004xm}
%\bibitem{Cheung:2004xm}
  K.~Cheung and O.~Seto,
  %``Phenomenology of TeV right-handed neutrino and the dark matter model,''
  Phys.\ Rev.\  D {\bf 69} (2004) 113009.
%  [arXiv:hep-ph/0403003].
  %%CITATION = PHRVA,D69,113009;%%

\bibitem{Ma}
%\cite{Ma:2006km}
%\bibitem{Ma:2006km}
  E.~Ma,
  %``Verifiable radiative seesaw mechanism of neutrino mass and dark matter,''
  Phys.\ Rev.\  D {\bf 73} (2006) 077301;
%  [arXiv:hep-ph/0601225].
  %%CITATION = PHRVA,D73,077301;%%
%
%\cite{Ma:2007gq}
%\bibitem{Ma:2007gq}
%  E.~Ma,
  %``Z_3 Dark Matter and Two-Loop Neutrino Mass,''
  Phys.\ Lett.\  B {\bf 662} (2008) 49;
%  [arXiv:0708.3371 [hep-ph]].
  %%CITATION = PHLTA,B662,49;%%
%
%\cite{Hambye:2006zn}
%\bibitem{Hambye:2006zn}
  T.~Hambye, K.~Kannike, E.~Ma and M.~Raidal,
  %``Emanations of Dark Matter: Muon Anomalous Magnetic Moment, Radiative
  %Neutrino Mass, and Novel Leptogenesis at the TeV Scale,''
  Phys.\ Rev.\  D {\bf 75} (2007) 095003;
%  [arXiv:hep-ph/0609228].
  %%CITATION = PHRVA,D75,095003;%%
%
%\cite{Ma:2008cu}
%\bibitem{Ma:2008cu}
  E.~Ma and D.~Suematsu,
  %``Fermion Triplet Dark Matter and Radiative Neutrino Mass,''
  Mod.\ Phys.\ Lett.\  A {\bf 24} (2009) 583.
%  [arXiv:0809.0942 [hep-ph]].
  %%CITATION = MPLAE,A24,583;%%

\bibitem{AKS}
%\cite{Aoki:2008av}
%\bibitem{Aoki:2008av}
  M.~Aoki, S.~Kanemura and O.~Seto,
  %``Neutrino mass, Dark Matter and Baryon Asymmetry via TeV-Scale Physics
  %without Fine-Tuning,''
  Phys.\ Rev.\ Lett.\  {\bf 102} (2009) 051805;
%  [arXiv:0807.0361 [hep-ph]].
  %%CITATION = PRLTA,102,051805;%%
%
%\cite{Aoki:2009vf}
%\bibitem{Aoki:2009vf}
%  M.~Aoki, S.~Kanemura and O.~Seto,
  %``A Model of TeV Scale Physics for Neutrino Mass, Dark Matter and Baryon
  %Asymmetry and its Phenomenology,''
  Phys.\ Rev.\  D {\bf 80} (2009) 033007;
%  [arXiv:0904.3829 [hep-ph]].
  %%CITATION = PHRVA,D80,033007;%%
%
%\cite{Aoki:2011zg}
%\bibitem{Aoki:2011zg}
  M.~Aoki, S.~Kanemura and K.~Yagyu,
  %``Triviality and vacuum stability bounds in the three-loop neutrino mass
  %model,''
  Phys.\ Rev.\  D {\bf 83} (2011) 075016;
%  [arXiv:1102.3412 [hep-ph]].
  %%CITATION = PHRVA,D83,075016;%%
%
%\cite{Aoki:2011yk}
%\bibitem{Aoki:2011yk}
%  M.~Aoki, S.~Kanemura and K.~Yagyu,
  %``Doubly-charged scalar bosons from the doublet,''
  Phys.\ Lett.\  B {\bf 702} (2011) 355.
%  [arXiv:1105.2075 [hep-ph]].
  %%CITATION = PHLTA,B702,355;%%
  
%\cite{Lyth:1998xn}
\bibitem{slow-roll} 
%%\cite{Linde:1981mu} 
%\bibitem{Linde:1981mu}
  A.~D.~Linde,
  %``A New Inflationary Universe Scenario: A Possible Solution of the Horizon, Flatness, Homogeneity, Isotropy and Primordial Monopole Problems,''
  Phys.\ Lett.\ B {\bf 108} (1982) 389;
  %%CITATION = PHLTA,B108,389;%%
%
%%\cite{Albrecht:1982wi}
%\bibitem{Albrecht:1982wi}
  A.~Albrecht and P.~J.~Steinhardt,
  %``Cosmology for Grand Unified Theories with Radiatively Induced Symmetry Breaking,''
  Phys.\ Rev.\ Lett.\  {\bf 48} (1982) 1220.
  %%CITATION = PRLTA,48,1220;%%

%\cite{Aprile:2012nq}
\bibitem{XENON100} 
  E.~Aprile {\it et al.}  [XENON100 Collaboration],
  %``Dark Matter Results from 225 Live Days of XENON100 Data,''  
  Phys.\ Rev.\ Lett.\  {\bf 109} (2012) 181301.  %[arXiv:1207.5988 [astro-ph.CO]].  %%CITATION = ARXIV:1207.5988;%%

%%%%%%%  ILC  %%%%%%%%%
\bibitem{ILC1}
%
%\cite{BrauJames:2007aa}
%\bibitem{BrauJames:2007aa} 
  J.~Brau, (Ed.) {\it et al.}  [ILC Collaboration],
  %``ILC Reference Design Report: ILC Global Design Effort and World Wide Study,''
  arXiv:0712.1950 [physics.acc-ph];
  %%CITATION = ARXIV:0712.1950;%%
%
%
%\cite{Djouadi:2007ik}
%\bibitem{Djouadi:2007ik} 
  G.~Aarons {\it et al.}  [ILC Collaboration],
  %``International Linear Collider Reference Design Report Volume 2: Physics At The Ilc,''
  arXiv:0709.1893 [hep-ph];
  %%CITATION = ARXIV:0709.1893;%%
%
%
%\cite{Phinney:2007gp}
%\bibitem{Phinney:2007gp} 
  N.~Phinney, N.~Toge and N.~Walker,
  %``LC Reference Design Report Volume 3 - Accelerator,''
  arXiv:0712.2361 [physics.acc-ph];
  %%CITATION = ARXIV:0712.2361;%%
%
%
%\cite{Behnke:2007gj}
%\bibitem{Behnke:2007gj} 
  T.~Behnke, (Ed.) {\it et al.}  [ILC Collaboration],
  %``ILC Reference Design Report Volume 4 - Detectors,''
  arXiv:0712.2356 [physics.ins-det];
  %%CITATION = ARXIV:0712.2356;%%
%
H.~Baer, {\it et al.} "Physics at the International Linear Collider", 
{\it Physics Chapter of the ILC Detailed Baseline Design Report}:
http://lcsim.org/papers/DBDPhysics.pdf.

%\cite{Cui:2009xq}
\bibitem{direct_Z} 
  Y.~Cui, D.~E.~Morrissey, D.~Poland and L.~Randall,
  %``Candidates for Inelastic Dark Matter,''  
  JHEP {\bf 0905} (2009) 076;  %[arXiv:0901.0557 [hep-ph]].  %%CITATION = ARXIV:0901.0557;%%
%\cite{Arina:2009um}
%\bibitem{Arina:2009um} 
  C.~Arina, F.~-S.~Ling and M.~H.~G.~Tytgat,
  %``IDM and iDM or The Inert Doublet Model and Inelastic Dark Matter,''  
  JCAP {\bf 0910} (2009) 018.  %[arXiv:0907.0430 [hep-ph]].  %%CITATION = ARXIV:0907.0430;%%
  
%\cite{Kashiwase:2012xd}
\bibitem{Kashiwase:2012xd}
  S.~Kashiwase and D.~Suematsu,
  %``Baryon number asymmetry and dark matter in the neutrino mass model with an inert doublet,''  
  Phys.\ Rev.\ D {\bf 86} (2012) 053001.  %[arXiv:1207.2594 [hep-ph]].  %%CITATION = ARXIV:1207.2594;%%

%\cite{LopezHonorez:2006gr}
\bibitem{LopezHonorez:2006gr}
  L.~Lopez Honorez, E.~Nezri, J.~F.~Oliver and M.~H.~G.~Tytgat,
  %``The Inert Doublet Model: An Archetype for Dark Matter,''  
  JCAP {\bf 0702} (2007) 028.%  [hep-ph/0612275].  %%CITATION = HEP-PH/0612275;%%  %172 citations counted in INSPIRE as of 11 Mar 2013

\bibitem{beta}
%\cite{Inoue:1979nn}
%\bibitem{Inoue:1979nn} 
  K.~Inoue, A.~Kakuto and Y.~Nakano,
  %``Perturbation Constraint On Particle Masses In The Weinberg-salam Model With Two Massless Higgs Doublets,''  
  Prog.\ Theor.\ Phys.\  {\bf 63} (1980) 234;  %%CITATION = PTPKA,63,234;%%
%\bibitem{Komatsu:1981xh} 
H.~Komatsu,
%``Behavior Of The Yukawa And The Quartic Scalar Couplings In Grand Unified Theories,''  
Prog.\ Theor.\ Phys.\  {\bf 67} (1982) 1177.  %%CITATION = PTPKA,67,1177;%%

\bibitem{extended_run} 
%\cite{Nie:1998yn}
%\bibitem{Nie:1998yn} 
  S.~Nie and M.~Sher,
  %``Vacuum stability bounds in the two Higgs doublet model,''  
  Phys.\ Lett.\ B {\bf 449} (1999) 89;  %[hep-ph/9811234].  %%CITATION = HEP-PH/9811234;%%
%
%\cite{Kanemura:1999xf}
%\bibitem{Kanemura:1999xf} 
  S.~Kanemura, T.~Kasai and Y.~Okada,
  %``Mass bounds of the lightest CP even Higgs boson in the two Higgs doublet model,''  
  Phys.\ Lett.\ B {\bf 471} (1999) 182.  %[hep-ph/9903289].  %%CITATION = HEP-PH/9903289;%%

\bibitem{LEP_direct}
%\cite{Abbiendi:2003sc}
%\bibitem{Abbiendi:2003sc} 
  G.~Abbiendi {\it et al.}  [OPAL Collaboration],
  %``Search for chargino and neutralino production at s**(1/2) = 192-GeV to 209 GeV at LEP,''  
  Eur.\ Phys.\ J.\ C {\bf 35} (2004) 1;  %[hep-ex/0401026].  %%CITATION = HEP-EX/0401026;%%
%\cite{Abbiendi:2003ji}
%\bibitem{Abbiendi:2003ji} 
%  G.~Abbiendi {\it et al.}  [OPAL Collaboration],
  %``Search for anomalous production of dilepton events with missing transverse momentum in e+ e- collisions at s**(1/2) = 183-Gev to 209-GeV,''  
  Eur.\ Phys.\ J.\ C {\bf 32} (2004) 453.  %[hep-ex/0309014].  %%CITATION = HEP-EX/0309014;%%

%\cite{Pierce:2007ut}
\bibitem{LEP_pm} 
  A.~Pierce and J.~Thaler,
  %``Natural Dark Matter from an Unnatural Higgs Boson and New Colored Particles at the TeV Scale,''  
  JHEP {\bf 0708} (2007) 026. %[hep-ph/0703056 [HEP-PH]].  %%CITATION = HEP-PH/0703056;%%

%\cite{Lundstrom:2008ai}
\bibitem{LEP_HApair} 
  E.~Lundstrom, M.~Gustafsson and J.~Edsjo,
  %``The Inert Doublet Model and LEP II Limits,''  
  Phys.\ Rev.\ D {\bf 79} (2009) 035013.  %[arXiv:0810.3924 [hep-ph]].  %%CITATION = ARXIV:0810.3924;%%

%\cite{Peskin:1990zt}
\bibitem{STdef} 
  M.~E.~Peskin and T.~Takeuchi,
  %``A New constraint on a strongly interacting Higgs sector,''  
  Phys.\ Rev.\ Lett.\  {\bf 65} (1990) 964;  %%CITATION = PRLTA,65,964;%%
%\cite{Peskin:1991sw}
%\bibitem{Peskin:1991sw} 
%  M.~E.~Peskin and T.~Takeuchi,
  %``Estimation of oblique electroweak corrections,''  
  Phys.\ Rev.\ D {\bf 46} (1992) 381.  %%CITATION = PHRVA,D46,381;%%

%\cite{Toussaint:1978zm}
\bibitem{ST1} 
  D.~Toussaint,
  %``Renormalization Effects From Superheavy Higgs Particles,''  
  Phys.\ Rev.\ D {\bf 18} (1978) 1626;  %%CITATION = PHRVA,D18,1626;%%
%\cite{Peskin:2001rw}
%\bibitem{Peskin:2001rw} 
  M.~E.~Peskin and J.~D.~Wells,
  %``How can a heavy Higgs boson be consistent with the precision electroweak measurements?,''  
  Phys.\ Rev.\ D {\bf 64} (2001) 093003.  %[hep-ph/0101342].  %%CITATION = HEP-PH/0101342;%%

%%\cite{Kanemura:2011sj}
\bibitem{ST2}
  S.~Kanemura, Y.~Okada, H.~Taniguchi and K.~Tsumura,
  %``Indirect bounds on heavy scalar masses of the two-Higgs-doublet model in light of recent Higgs boson searches,''  
  Phys.\ Lett.\ B {\bf 704} (2011) 303;  %[arXiv:1108.3297 [hep-ph]].  %%CITATION = ARXIV:1108.3297;%%
%\cite{Baak:2011ze}
%\bibitem{Baak:2011ze} 
  M.~Baak, M.~Goebel, J.~Haller, A.~Hoecker, D.~Ludwig, K.~Moenig, M.~Schott and J.~Stelzer,
  %``Updated Status of the Global Electroweak Fit and Constraints on New Physics,''  
  Eur.\ Phys.\ J.\ C {\bf 72} (2012) 2003.  %[arXiv:1107.0975 [hep-ph]].  %%CITATION = ARXIV:1107.0975;%%

%\cite{Barbieri:2006dq}
\bibitem{LHC1}
  R.~Barbieri, L.~J.~Hall and V.~S.~Rychkov,
  %``Improved naturalness with a heavy Higgs: An Alternative road to LHC physics,''  
  Phys.\ Rev.\ D {\bf 74} (2006) 015007  [hep-ph/0603188].  %%CITATION = HEP-PH/0603188;%%  %313 citations counted in INSPIRE as of 11 Mar 2013

%\cite{Cao:2007rm}
\bibitem{LHC2} 
  Q.~-H.~Cao, E.~Ma and G.~Rajasekaran,
  %``Observing the Dark Scalar Doublet and its Impact on the Standard-Model Higgs Boson at Colliders,''  
  Phys.\ Rev.\ D {\bf 76} (2007) 095011.  %[arXiv:0708.2939 [hep-ph]].  %%CITATION = ARXIV:0708.2939;%%

%\cite{Dolle:2009ft}
\bibitem{LHC3}
  E.~Dolle, X.~Miao, S.~Su and B.~Thomas,
  %``Dilepton Signals in the Inert Doublet Model,''  
  Phys.\ Rev.\ D {\bf 81} (2010) 035003  %[arXiv:0909.3094 [hep-ph]].  %%CITATION = ARXIV:0909.3094;%%  %35 citations counted in INSPIRE as of 11 Mar 2013

%\cite{Pukhov:2004ca}
\bibitem{calc}
  A.~Pukhov,
  %``CalcHEP 2.3: MSSM, structure functions, event generation, batchs, and generation of matrix elements for other packages,''  
  hep-ph/0412191.  %%CITATION = HEP-PH/0412191;%%

\bibitem{ILC2}
%\cite{Aoki:2013lhm}
%\bibitem{Aoki:2013lhm}
  M.~Aoki, S.~Kanemura and H.~Yokoya,
  %``Reconstruction of Inert Doublet Scalars at the International Linear Collider,''
  arXiv:1303.6191 [hep-ph]; 
  %%CITATION = ARXIV:1303.6191;%%
  %
%\cite{Aoki:2010mu}
%\bibitem{inILC} 
  M.~Aoki and S.~Kanemura,
  %``Probing the Majorana nature of TeV-scale radiative seesaw models at collider experiments,''  
  Phys.\ Lett.\ B {\bf 689} (2010) 28.  %[arXiv:1001.0092 [hep-ph]].  %%CITATION = ARXIV:1001.0092;%%

\end{thebibliography}

\end{document}